\documentclass[reqno]{amsart}
\usepackage{url}
\usepackage{empheq}
\usepackage{verbatim}
\usepackage{alltt}
\usepackage{endnotes}
\usepackage[numbers]{natbib}
\usepackage{palatino}
\usepackage{mathpazo}
\usepackage{graphicx}
\graphicspath{{figures/}}
\usepackage{subfigure}
\graphicspath{{figures/}}
\usepackage{psfrag}
\usepackage{amsmath,amsthm,amscd,amssymb,amsbsy} 
\usepackage{latexsym}
\usepackage{sistyle}

\theoremstyle{definition}
\newtheorem{definition}{Definition}[section]
\newtheorem{example}{Example}[section]
\usepackage{hyperref}
\usepackage{algorithm}
\usepackage{algorithmic}
\usepackage{threeparttable}
\usepackage{Sweave}
\usepackage{fancyhdr}
\begin{document}
\title{Bernoulli Runs
\\
Using ``Book Cricket'' to Evaluate Cricketers}
\author{Anand Ramalingam}
\address{}
\email{anandram@gmail.com}
\begin{abstract} 
This paper proposes a simple method to evaluate batsmen and bowlers in cricket.
The idea in this paper refines ``book cricket'' and evaluates a batsman by
answering the question: How many runs a team consisting of same player 
replicated eleven times will score?
\subsection*{Keywords} 
  Probability, Combinatorics, Coin Toss, 
  Bernoulli Runs, Book Cricket, Monte Carlo Simulation.
\end{abstract} 
\maketitle
\section{Introduction}
In the late 1980s and early 1990s to beat afternoon drowsiness in school one resorted to 
playing ``book cricket''.  The book cricket rules was quite simple.  
Pick a text book and open it randomly and note the last digit of the even numbered page. 
The special case is when you see a page ending with $0$ then you have lost a wicket. 
If you see $8$, most of my friends would score it as a $1$ run. 
The other digits {$2,4,6$} would be scored as the same, that is if you see
a page ending with $4$ then you have scored $4$ runs.
We would play two national teams without any overs limit since it was a 
too much of a hassle to count the number of deliveries\footnote{The number of 
deliveries equals the number of times we opened the book.}. 

A book cricketer would construct his own team and match it up against his friend. 
One of the teams would be the Indian team and the other team 
would be the side the Indian team was playing at that time.
The book cricketer would play till he lost $10$ wickets\footnote{Remember losing a wicket is 
when on opening the book, you see a page number that ends with a $0$.  
Thus the inning ends when he sees ten page numbers that end with a $0$.}.  
Thus it was like test cricket but with just one innings for each team.
The one who scores the most number of runs would be declared the winner.

In probabilistic terms, we were simulating a batsman with the following
probability mass function:
\begin{align}
\label{eqn:book:cricket:prob:model}
p_{X}(x) &=  \frac{1}{5} \quad  x \in \{\mathrm{out},1,2,4,6\}
\end{align}

With this simple model, we used to get weird results like a well-known batting
bunny like Narendra Hirwani scoring the most number of runs. So we had to change
rules for each player based on his batting ability. For example, a player like
Praveen Amre would be dismissed only if we got consecutive page numbers that ended
with a $0$. Intuitively without understanding a whole lot of probability, we had
reduced the probability of Amre being dismissed from $\frac{1}{5}$ to $\frac{1}{25}$. 
In the same spirit, for Hirwani we modified the original model and made $8$ a
dismissal. Thus the probability of Hirwani being dismissed went up from 
$\frac{1}{5}$ to $\frac{2}{5}$ thus reducing the chances of Hirwani getting 
the highest score which did not please many people in the class\footnote{ 
The another reason for the lack of popularity was that it involved more book keeping.}. 

In the next section, we develop a probabilistic model of cricket by refining
the book cricket model. The refinement is in probabilities which is updated to 
approximate the real world cricketing statistics. 
\section{A Probabilistic Model for Cricket}
The book cricket model described above can be thought of as a five-sided die game.
The biggest drawback of the book cricket model described in 
Eq.~\eqref{eqn:book:cricket:prob:model} is the fact all the five outcomes,  
\begin{align}
\Omega &= \{\mathrm{out},1,2,4,6\} \notag
\end{align}
are equally likely.

Instead of assigning uniform probabilities to all the five outcomes, a realistic
model will take probabilities from the career record. 
For example, 
\href{http://www.espncricinfo.com/india/content/player/30750.html}
{VVS Laxman}
has hit $4$ sixes in $3282$ balls he faced in his ODI career.
Thus one can model the probability of Laxman hitting a six as $\frac{4}{3282}$
instead of our naive book cricket probability of $\frac{1}{5}$.

To further refine this model, we have to add two more outcomes to the model namely:
the dot ball (resulting in $0$ runs) and the scoring shot which results in $3$ runs 
being scored. This results in a seven-sided die model whose sample space is:
\begin{align}
\Omega &= \{\mathrm{out},0,1,2,3,4,6\} \label{eqn:seven-sided}
\end{align}
\subsection{A Simplified Model}
We need to simplify our model since it is hard to obtain detailed statistics 
for the seven-sided die model. 
For example, it is pretty hard to find out how many twos and threes Inzamam-ul-Haq 
scored in his career.  But it is easy to obtain his career average and strike rate.

\begin{definition}
The \textbf{average} ($\mathrm{avg}$) for a \emph{batsman} is defined as the average number of runs scored
per dismissal.
    \begin{align}
    \mathrm{avg} &= \frac{\text{number of runs scored}} {\text{number of dismissals}} \label{eqn:def:avg}
    \end{align}

In case of a \emph{bowler} the numerator becomes number of runs \emph{conceded}.
\end{definition}

\begin{definition}
The \textbf{strike rate} ($\mathrm{sr}$) for a batsman is defined as the runs scored per $100$ balls. 
Thus when you divide the strike rate by $100$ you get the average runs ($r$) scored per ball.
    \begin{align}
    \mathrm{r} = \frac{\mathrm{sr}}{100}
    &= \frac {\text{number of runs scored}} {\text{number of balls faced}} \label{eqn:def:batsman:r}
    \end{align}
\end{definition}

\begin{definition}
The \textbf{economy rate} ($\mathrm{econ}$) for a bowler is defined as the runs 
conceded per $6$ balls\footnote{ 
In the case of bowlers, strike rate ($\mathrm{sr}$) means number of deliveries
required on a average for a dismissal. 
So strike rate ($\mathrm{sr}$) is interpreted differently for a batsman and bowler. 
An example of context sensitive information or in Object Oriented jargon overloading!
}.
Thus the average runs ($r$) conceded per ball is:
    \begin{align}
    \mathrm{r} = \frac{\mathrm{econ}}{6}
    &= \frac {\text{number of runs conceded}} {\text{number of balls bowled}} \label{eqn:def:bowler:r}
    \end{align}
\end{definition}

We need to simplify the seven-sided die probabilistic model to take advantage
of the easy availability of the above statistics: average, strike rate and
economy rate.
This simplification leads to modeling of every delivery in cricket 
as simple coin tossing experiment. 

Thus the simplified model will have only two outcomes, either a \emph{scoring shot} (heads) 
or a \emph{dismissal} (tails) during every delivery. 
We also assign a value to the scoring shot (heads) namely the average runs scored per ball. 

Now to complete our probabilistic model all we need is to estimate the probability of getting 
dismissed\footnote{The probability of getting a tail.}.
From the definitions of strike rate and average, one can calculate the 
average number of deliveries a batsman faces before he is dismissed ($\mathrm{bpw}$). 
\begin{align}
\mathrm{bpw} &= \frac{\text{number of balls faced}} {\text{number of dismissals}} \notag \\
    &= \frac{\text{number of runs scored}} {\text{number of dismissals}} \times
       \frac{\text{number of balls faced}} {\text{number of runs scored}} \notag \\
       &= \mathrm{avg} \times \frac{1}{\frac{\mathrm{sr}}{100}} \notag \\
       &= 100 \times \left( \frac{\mathrm{avg}} {\mathrm{sr}} \right) \label{eqn:bpw}
\end{align}

Assuming that the batsman can be dismissed during any delivery,
the probability of being dismissed ($1-p$) is given by\footnote{
The formula 
  $\mathbb{P}\{\text{Dismissal}\} = 1-p = \frac{r} {\mathrm{avg}}$
also applies to a bowler with the $r$ being calculated using 
Eq.~\eqref{eqn:def:bowler:r}.
}:
\begin{align}
  \mathbb{P}\{\text{Dismissal}\} = 1-p  &= \frac{1}{\text{bpw}} 
          = \frac{\mathrm{sr}} {100 \times \mathrm{avg}}
          = \frac{r} {\mathrm{avg}} \label{eqn:batsman:prob:out}
\end{align}

In probability parlance, coin tossing is called a Bernoulli 
trial~\cite{ross:2002:probability}.
From the perspective of a batsman, if you get a head you score $r$ runs
and if you get a tail you are dismissed.
Thus the most basic event in cricket, the ball delivered by a bowler to a
batsman is modeled by a coin toss.

Just as Markov chains form the theoretical underpinning for modeling baseball 
run scoring~\cite{dangelo:2010:baseball}, Bernoulli\footnote{
Arguably, the Bernoulli's were the greatest mathematical family that ever 
lived~\cite{mukhopadhyay:2001:bernoulli,polasek:2000:bernoulli}
} 
trials form the basis 
for cricket. 
Now that we have modeled each delivery as a Bernoulli trial, we now have the
mathematical tools to evaluate a batsman or bowler.

\section{Evaluating a Batsman}
To evaluate a batsman we imagine a ``team'' consisting of eleven replicas of the same 
batsman and find how many \emph{runs on average} this imaginary team will 
score\footnote{
It is straightforward to apply this method to evaluate a bowler too.
}.
For example, to evaluate Sachin Tendulkar we want to find out 
how many runs will be scored by a team consisting of eleven Tendulkar's.

Since we model each delivery as a Bernoulli trial, the total runs scored by this
imaginary team will be a probability distribution. 
To further elaborate this point, if a team of Tendulkar's faces $300$ balls, 
they score $300r$ runs if they don't lose a wicket where $r$ is defined in 
Eq.~\eqref{eqn:def:batsman:r}.
They score $299r$ runs if they lose only one wicket.  On the other end of the scale,
this imaginary team might be dismissed without scoring a run if it so happens 
that all the first $10$ tosses turn out to be tails\footnote{
The probability of  being all out without a run being scored will be astronomically 
low for an imaginary team of eleven Tendulkar's!}. 
Thus this imaginary team can score total runs anywhere between $[0, 300r]$ 
and each total is associated with probability.

But it is difficult to interpret probability distribution and it is much 
easier to comprehend basic statistical summaries such as \emph{mean} 
and \emph{standard deviation}.
We call the \emph{mean} as \textbf{Bernoulli runs} in this paper since 
the idea was inspired by Markov runs in Baseball~\cite{dangelo:2010:baseball}.

\subsection{Bernoulli Runs}
We now derive the formula for the \emph{mean} of runs scored by a team
consisting of eleven replicas of the same batsman in an One day international
(ODI) match\footnote{The ODI has a maximum of $300$ deliveries per team.
The formulas can be derived for Twenty20 and Test matches with appropriate
deliveries limit.}.

Let $Y$ denote the number of runs scored in a ODI by this imaginary team.
It is easier to derive the formula for mean of the total runs ($\mathbb{E}(Y)$) 
scored by partitioning the various scenarios into two cases:
\begin{enumerate}
\item The team loses all the wickets ($\mathbb{E}_{\text{all-out}}(Y)$);
\item The team uses up all the allotted deliveries which implies that the team
has lost less than 10 wickets in the allotted deliveries ($\mathbb{E}_{\overline{\text{all-out}}}(Y)$).
\end{enumerate}

This leads to:
\begin{align}
\mathbb{E}(Y) &= \mathbb{E}_{\text{all-out}}(Y) + \mathbb{E}_{\overline{\text{all-out}}}(Y)
                  \label{eqn:bruns:split}
\end{align}

In the first case of team losing all the wickets, we can once again partition 
on the delivery the tenth wicket was lost.
Let $b$ be the delivery the tenth wicket fell. The tenth wicket can fall on any
delivery between $[10,300]$.
The first nine wickets could have fallen in any one of the previous $b-1$ deliveries.
The number of possible ways the nine wickets could have fallen in $b-1$
deliveries is given by ${b-1 \choose 9}$. 
The number of scoring shots (heads in coin tosses) is $b-10$.
The mean number of runs scored while losing all wickets is given by: 
\begin{align}
\mathbb{E}_{\text{all-out}}(Y) &= r \sum_{b=10}^{300} (b-10) 
                                                        \left( 
                                                          {b-1 \choose 9} p^{(b-1)-9} (1-p)^9 
                                                        \right)
                                                        (1-p) 
                                                      \label{eqn:bruns:all-out}
\end{align}

The second case can be partitioned on basis of the number of wickets ($w$) lost.
Applying the same logic, one can derive the following result for the mean
number of runs scored:
\begin{align} 
\mathbb{E}_{\overline{\text{all-out}}}(Y) &=
              r \sum_{w=0}^{9} (300-w) {300 \choose w} p^{300-w} (1-p)^w
                                                      \label{eqn:bruns:not-all-out}
\end{align}

Substituting Eq.~\eqref{eqn:bruns:all-out} and Eq.~\eqref{eqn:bruns:not-all-out} in
Eq.~\eqref{eqn:bruns:split} we get the following equation for the mean of the runs
scored:
\begin{multline}
\mathbb{E}(Y) = r \sum_{b=10}^{300} (b-10) {b-1 \choose 9} p^{b-10} (1-p)^{10} \\
                   + r \sum_{w=0}^{9} (300-w) {300 \choose w} p^{300-w} (1-p)^w
                                                      \label{eqn:bruns:mean}
\end{multline}

One can generalize the above Eq.~\eqref{eqn:bruns:mean} to generate any moment. 
The $k$th moment is given by:
\begin{multline}
\mathbb{E}(Y^{k}) = r^{k} \sum_{b=10}^{300} (b-10)^{k} {b-1 \choose 9} p^{b-10} (1-p)^{10} \\
                   + r^{k} \sum_{w=0}^{9} (300-w)^{k} {300 \choose w} p^{300-w} (1-p)^w
                                                      \label{eqn:bruns:moment}
\end{multline}

The standard deviation can be obtained by 
\begin{align}
  \sigma_{Y} &= \sqrt{\mathbb{E}(Y^{2}) - \left( \mathbb{E}(Y) \right)^{2}}
                                                      \label{eqn:bruns:sd}
\end{align}

To make things concrete, we illustrate the calculation of Bernoulli runs 
for a batsman and a bowler using the statistical programming language 
\verb@R@~\cite{R:manual}. The \verb@R@ code which implements this is 
listed in Appendix~\ref{appendix:formula}.
\begin{example}[Batsman]
\href{http://www.espncricinfo.com/westindies/content/player/52812.html}
{Sir Viv Richards}. 
Richards has an $\mathrm{avg} = 47.00$ and $\mathrm{sr} = 90.20$ in ODI matches.
From Eq.~\eqref{eqn:def:batsman:r} we get $r = \frac{\mathrm{sr}}{100} = 0.9020$ and 
from Eq.~\eqref{eqn:batsman:prob:out}, we get $1-p = \frac{0.9020}{47.00} = 0.01919$ 

Substituting the values of $1-p$ and $r$ in Eq.~\eqref{eqn:bruns:mean} and
Eq.~\eqref{eqn:bruns:sd} we get $\mathrm{mean} = 262.84$ and 
$\mathrm{sd} = 13.75$. One can interpret the result as, a team consisting of
eleven Richards' will score on average $262.84$ runs per ODI inning with a 
standard deviation of $13.75$ runs per inning. 
The code listed in Appendix~\ref{appendix:formula} is at \verb@R/analytical.R@
and can be executed as follows:
\begin{Schunk}
\begin{Sinput}
> source(file = "R/analytical.R")
> bernoulli(avg = 47, sr = 90.2)
\end{Sinput}
\begin{Soutput}
$mean
[1] 262.8434

$sd
[1] 13.75331
\end{Soutput}
\end{Schunk}
\end{example}

\begin{example}[Bowler]
\href{http://www.espncricinfo.com/westindies/content/player/51107.html}
{Curtly Ambrose}. 
Ambrose
has an $\mathrm{avg} = 24.12$ and $\mathrm{econ} = 3.48$ in ODI matches.
From Eq.~\eqref{eqn:def:bowler:r}, we get $r = \frac{\mathrm{econ}}{6} = 0.58$ and
from Eq.~\eqref{eqn:batsman:prob:out}, we get $1-p = \frac{r}{\mathrm{avg}} = \frac{0.58}{24.12} = 0.024$

Substituting the values of $1-p$ and $r$ in Eq.~\eqref{eqn:bruns:mean} and
Eq.~\eqref{eqn:bruns:sd} we get $\mathrm{mean} = 164.39$ and 
$\mathrm{sd} = 15.84$. One can interpret the result as, a team consisting of
eleven Ambrose's will concede on average $164.39$ runs per ODI inning with a 
standard deviation of $15.84$ runs per inning.

\begin{Schunk}
\begin{Sinput}
> bernoulli(avg = 24.12, sr = 3.48 * 100/6)
\end{Sinput}
\begin{Soutput}
$mean
[1] 164.3869

$sd
[1] 15.84270
\end{Soutput}
\end{Schunk}
\end{example}

\subsubsection{Poisson process} 
An aside. One can also use Poisson process to model a batsman's career.
This is because the probability of getting dismissed is
pretty small ($q \rightarrow 0$), and the number of deliveries a player faces 
is pretty high over his entire career ($n$).
Poisson distribution can used to model the rare events (dismissal) 
counting with parameter $\lambda = nq$~\cite{ross:2002:probability}. 
The $\lambda$ can be interpreted as the average number of wickets 
that a team will lose in $n$ balls. For example, a team of eleven Richards
will lose $300 \times 0.01919 = 5.78$ wickets on an average which explains
the reason why his standard deviation is very low.
\subsubsection{Monte Carlo Simulation}
Another aside. The Monte Carlo simulation code for the probability model 
proposed in this paper is listed in Appendix~\ref{appendix:montecarlo}  
in \verb@R@. 
Monte Carlo simulation can be used to verify the formula for Bernoulli runs
we have derived. 
In other words, it provides another way find the Bernoulli runs.

Also as one refines the model it becomes difficult to obtain a closed form 
solution to the Bernoulli runs and Monte Carlo simulation comes in handy
during such situations. 
For example, it is straightforward to modify the code 
to generate Bernoulli runs using the seven-sided die model presented in 
Eq.~\eqref{eqn:seven-sided}. Thus any model can be simulated using 
Monte Carlo.
\section{Reward to Risk Ratio} 
\href{http://www.espncricinfo.com/india/content/player/35263.html}
{Virender Sehwag}
has an $\mathrm{avg} = 34.64$ and $\mathrm{sr} = 103.27$ in ODI 
matches\footnote{
The statistics for current players are up to date as of December 31, 2010.
}
this leads to Bernoulli runs (mean) = $275.96$ and 
standard deviation = $42.99$. Thus on an average, Sehwag scores more runs
than Richards but he is also risky compared to Richards.  
To quantify this, we borrow the concept of 
\href{http://en.wikipedia.org/wiki/Sharpe_ratio}
{Sharpe Ratio} from the world of Financial Mathematics and we call it 
Reward to Risk Ratio ($\mathrm{RRR}$). 
\begin{definition}
The Reward to Risk Ratio ($\mathrm{RRR}$) for a \emph{batsman} is defined as:
\begin{align}
\mathrm{RRR}  &= \frac{\mathbb{E}(Y) - c_{\text{batsman}}} {\sigma_{Y}} 
  \label{eqn:batsman:rrr}
\end{align}
and for a \emph{bowler} it is defined as:
\begin{align}
\mathrm{RRR}  &= \frac{c_{\text{bowler}} - \mathbb{E}(Y)} {\sigma_{Y}}
  \label{eqn:bowler:rrr}
\end{align}
where  $\mathbb{E}(Y)$ is defined in Eq.~\eqref{eqn:bruns:mean} and
$\sigma_{Y}$ is defined in Eq.~\eqref{eqn:bruns:sd}. The constants 
$c_{\text{batsman}}$ and $c_{\text{bowler}}$ are discussed below.
\end{definition}
\subsection{Constants}
The Duckworth-Lewis (D/L) method predicts an 
\href{http://static.espncricinfo.com/db/ABOUT_CRICKET/RAIN_RULES/DL_FAQ.html}
{average score}\footnote{The pertinent question is No.13 in the 
hyperlinked D/L FAQ.} of $235$ runs will be scored by a team in an ODI match.
Though D/L average score seems to be a good candidate for usage as the constant
in $\mathrm{RRR}$, we use scale it before using it. The reason for scaling
is due to a concept named Value over Replacement player (VORP) which
comes from Baseball~\cite{wiki:vorp}.
Replacement player is a player who plays at the next rung below international 
cricket\footnote{ In India, the replacement players play in Ranji Trophy.}.
A team full of replacement players will have no risk and hence no upside.
The baseball statisticians have set the scale factor for replacement players
to be $20\%$ worse than the international players. Thus for batsman 
\begin{align}
c_{\text{batsman}} &= 0.8 \times 235 = 188 \label{eqn:batsman:c}
\end{align}
and for a bowler it is 
\begin{align}
c_{\text{bowler}} &= 1.2 \times 235 = 282 \label{eqn:bowler:c}
\end{align}
Thus a team of replacement batsman will end up scoring $188$ runs while 
a team of replacement bowlers will concede $282$ runs.
We end this paper by listing Bernoulli runs, standard deviation and reward to
risk ratio for some of the Indian ODI cricketers of 2010.
\begin{table*}[hbt]
\caption{Bernoulli Runs for batsmen}
\label{tbl:batsmen}
\begin{center}
\begin{tabular}{|l|r|r|r|r|r|}
\hline
\multicolumn{1}{|c|}{Name} &
\multicolumn{1}{|c|}{$\mathrm{avg}$} &
\multicolumn{1}{|c|}{$\mathrm{sr}$} &
\multicolumn{1}{|c|}{mean} &
\multicolumn{1}{|c|}{sd} &
\multicolumn{1}{|c|}{$\mathrm{RRR}$} 
\\\hline\hline
Virender Sehwag & 34.64 & 103.27 & 275.96 & 42.99 & 2.05\\\hline
Sachin Tendulkar & 45.12 & 86.26 & 251.43 & 13.01 & 4.88\\\hline
Gautam Gambhir & 40.43 & 86.52 & 249.52 & 17.75 & 3.47\\\hline
Yuvraj Singh & 37.06 & 87.94 & 249.84 & 23.29 & 2.66\\\hline
Mahendra Singh Dhoni & 50.28 & 88.34 & 258.87 & 10.42 & 6.80\\\hline
Suresh Raina & 36.11 & 90.15 & 253.62 & 26.79 & 2.45\\\hline
Yusuf Pathan & 29.33 & 110.00 & 261.15 & 59.59 & 1.23\\\hline
\end{tabular}
\end{center}
\end{table*}
\begin{table*}[hbt]
\caption{Bernoulli Runs for bowlers}
\label{tbl:bowlers}
\begin{center}
\begin{tabular}{|l|r|r|r|r|r|}
\hline
\multicolumn{1}{|c|}{Name} &
\multicolumn{1}{|c|}{$\mathrm{avg}$} &
\multicolumn{1}{|c|}{$\mathrm{econ}$} &
\multicolumn{1}{|c|}{mean} &
\multicolumn{1}{|c|}{sd} &
\multicolumn{1}{|c|}{$\mathrm{RRR}$} 
\\\hline\hline
Zaheer Khan & 29.85 & 4.91 & 224.89 & 29.44 & 1.94\\\hline
Praveen Kumar & 33.57 & 5.07 & 237.30 & 25.56 & 1.75\\\hline
Ashish Nehra & 31.03 & 5.15 & 235.25 & 31.40 & 1.49 \\\hline
Harbhajan Singh & 32.84 & 4.30 & 206.19 & 15.46 & 4.90\\\hline
Yusuf Pathan & 34.06 & 5.66 & 258.45 & 34.59 & 0.68 \\\hline
Yuvraj Singh & 39.76 & 5.04 & 242.61 & 16.66 & 2.36\\\hline
\end{tabular}
\end{center}
\end{table*}

It is clear from Table~\ref{tbl:bowlers} it is quite unfair to compare 
Zaheer Khan with Harbhajan Singh. Zaheer operates usually in the manic 
periods of power plays and slog overs while Harbhajan bowls mainly in the 
middle overs. But until we get detailed statistics the adjustments that go 
with it have to wait. 
\section*{Acknowledgments}
The statistics for the cricketers in this paper were taken from the 
\href{http://www.espncricinfo.com}{\texttt{ESPNcricinfo}} website.
The paper was typeset using \LaTeX{} and calculations were performed using
\href{http://www.r-project.org}{\texttt{R}} statistical computing system.
The author thanks countless contributors who made the above software in 
particular and open source software in general.
\bibliographystyle{plain}
\bibliography{report}
\normalsize
\appendix
\newpage
\section{Combinatorial Formula}
\label{appendix:formula}
\begin{verbatim}
#---------------------------------------------------------
# Generate Bernoulli runs using Combinatorial formula
# (*) Example - Batsman - Sir Viv Richards
# > bernoulli(avg=47, sr=90.2);
# (*) Example - Bowler - Curtly Ambrose
# > bernoulli(avg=24.12, sr=3.48*100/6);
#---------------------------------------------------------
bernoulli <- function(avg, sr, wickets = 10, balls = 300) {
# probability of scoring `r' runs in a given ball
  r <- sr/100;
# probability of dismissal
  q <- r/avg; p <- 1-q;

# runs scored while losing all the wickets
# you can lose 10 wickets in 10 balls or .... or in 300 balls
# all these events are independent hence you can add them
  b <- wickets:balls;
  k <- 1; Eallout  <- moments.allout(p, r, b, k, wickets);
  k <- 2; Eallout2 <- moments.allout(p, r, b, k, wickets);

# calculate the runs scored if less than 10 wickets fall in 
# the alloted number of balls
  w <- 0:wickets-1;
  k <- 1; Enot.allout  <- moments.not.allout(p, r, w, k, balls);
  k <- 2; Enot.allout2 <- moments.not.allout(p, r, w, k, balls);

# mean (bernoulli runs)
  ebr <- Eallout + Enot.allout;
# standard deviation
  ebr2 <- Eallout2 + Enot.allout2; sdbr <- sqrt(ebr2 - (ebr)^2);

  result <- list(mean=ebr, sd=sdbr);
  return(result);
}

moments.allout <- function(p, r, b, k, w = 10) {
  y <- ((b-w)^k)*choose(b-1,w-1)*p^(b-w)*(1-p)^w;
  eyk <- (r^k)*(sum(y));
  return(eyk);
}

moments.not.allout <- function(p, r, w, k, n = 300) {
  y <- ((n-w)^k)*choose(n,w)*p^(n-w)*(1-p)^w;
  eyk <- (r^k)*(sum(y));
  return(eyk);
}
#---------------------------------------------------------
\end{verbatim}

\newpage
\section{Monte Carlo Simulation}
\label{appendix:montecarlo}
\begin{verbatim}
#---------------------------------------------------------
# Bernoulli runs using Monte-Carlo simulation
#---------------------------------------------------------
bernoulli.monte.carlo <- function(avg, sr, simulations = 1000) {
# probability of scoring a run `r' in a given ball
  r <- sr/100;
# probability of dismissal
  q <- r/avg; p <- 1-q;

# runs scored = (avg runs per ball)*(number of scoring shots)
  runs <- r*sapply(1:simulations, function(x) simulate.inning(p));

  result <- list(mean=mean(runs), sd=sd(runs));
  return(result);
}

#---------------------------------------------------------
# Simulate an inning as if the same batsman plays
# every delivery till he faces max deliveries (300 in ODI)
# or till he gets out 10 times whichever is earlier.
#---------------------------------------------------------
simulate.inning <- function(p, balls = 300, wickets = 10) {
# toss the coin `n' times 
# tail - out; head - scoring shot 
  result <- rbinom(balls, 1, p);

# find the deliveries in which a wicket fell 
  fall.of.wicket <- which(!result);

# find the number of heads (scoring shots) 
# till the fall of the last wicket
  nheads <- 0;
  if (length(fall.of.wicket) < wickets) {
# team has not been bowled out
    nheads <- sum(result);
  } else {
# team has been bowled out
    last.wicket.index = fall.of.wicket[wickets];
    nheads <- sum(result[1:last.wicket.index]);
  }

  return(nheads);
}
#---------------------------------------------------------
\end{verbatim}
\end{document}